\documentclass[twocolumn,showpacs,nofootinbib,amsmath,amssymb,prd]{revtex4}

\usepackage{hyperref}
\usepackage{amsfonts}
\usepackage{graphicx}

\begin{document}

\title{Quasi-circular orbits of conformal thin-sandwich puncture binary black holes}

\author{Mark D. Hannam}
\email{marko@phys.utb.edu}
\affiliation{Center for Gravitational Wave Astronomy, University of Texas at Brownsville, 80 Fort Brown, Brownsville, TX 78520}

\begin{abstract} 
I construct initial data for equal-mass irrotational binary black holes using the conformal thin-sandwich puncture (CTSP) approach. I locate quasi-circular orbits using the effective-potential method, and estimate the location of the innermost stable circular orbit (ISCO). The ISCO prediction is consistent with results for conformal thin-sandwich data produced using excision techniques. These results also show that the ISCOs predicted by the effective-potential and ADM-Komar mass-comparison methods agree for conformal thin-sandwich data, just as they did for Bowen-York data. 
\end{abstract}


\maketitle

\section{Introduction}

One of the most fervently pursued goals of numerical relativity is the simulation of the final orbits, plunge and merger of two black holes, and an estimation of the gravitational radiation emitted in the process. The strongest gravitational-wave signal is expected to be emitted during the plunge. As such, the most useful numerical simulations will describe the binary-black-hole system from its last orbit through to the ringdown of the single final black hole. It is now possible to evolve binary-black-hole spacetimes on an orbital timescale \cite{brugmann04,alcubierre05}, and to extract gravitational waveforms from evolutions with even modest survival times \cite{baker02a}. If these waveforms are to be astrophysically realistic, the initial data for the numerical simulations must also be realistic. As such, two criteria for useful initial data are that (I) they describe two astrophysically realistic black holes in orbit, and (II) this should be the last, or almost the last, orbit before the black holes plunge together. Due to the circularizing effect of gravitational-radiation emission during the black holes' inspiral, we expect these orbits to be almost circular.

The initial data used for all of the numerical evolutions cited above were those based on the Bowen-York solution of the momentum constraint in the conformal transverse-traceless (CTT) decomposition of the initial-value equations \cite{york79,bowen80}. These data are suspect on both of the above criteria for good initial data. They are known to be astrophysically unrealistic: they are conformally flat, which is correct only to second post-Newtonian order even for single spinning black holes, and they contain spurious gravitational radiation \cite{cook00}. In addition, predictions of the innermost stable circular orbit (ISCO) for these data disagree with third-order post-Newtonian (3PN) predictions by as much as a factor of two \cite{damour02}, and numerical evolutions of Bowen-York black holes that were predicted to be in quasi-circular orbit, are instead found to immediately plunge together \cite{baker02a,alcubierre05}. If we are to obtain meaningful results for numerical evolutions, we require more astrophysically realistic initial data, and a better estimate of the ISCO.

A promising alternative to the CTT decomposition is York's conformal thin-sandwich (CTS) formulation \cite{york98,york03}. In all formulations of the initial-value equations, there are some pieces of data that we are free to specify as we wish. In the CTS formulation, the free data are more closely linked to the dynamics of the spacetime than in the CTT decomposition, and thus allow us to make physically motivated choices. 

The CTS equations have been solved for binary-black-hole initial data by a number of groups \cite{ggba,ggbb,pfeiffer02,cook04,yo04,ansorg05}, and all estimates of the ISCO using CTS data are in far better agreement with post-Newtonian results \cite{ggbb,cook04,yo04} than their Bowen-York counterparts. There are no published results on the evolution of CTS data, but when Bowen-York data are evolved from an initial separation greater than the CTS ISCO prediction, they are found to orbit \cite{brugmann04}, further supporting the CTS ISCO over the Bowen-York-data prediction. 

All CTS black-hole solutions to date were found using excision techniques: a region around each black hole is excised from the computational domain, and inner boundary conditions are placed on the resulting excision surface. The introduction of excision surfaces is numerically complex, and one is then presented with the problem of choosing appropriate inner boundary conditions, of which there are several choices \cite{ggba,cook01,cook04,yo04}. A simpler approach is to use punctures, introduced by Brandt and Br\"{u}gmann \cite{brandt97} to solve the Hamiltonian constraint for Bowen-York data. In the puncture approach, divergences at the black-hole locations (``punctures'') are removed analytically, and the resulting equation(s) can be solved numerically over the entire computational domain; no regions need be excised. The puncture approach is far easier to implement than excision, and has proven useful in constructing Bowen-York binary-black-hole initial data \cite{brandt97,baum00,ansorg04}, and in numerical evolutions \cite{alcubierre03,brugmann04,imbiriba04,alcubierre05,alcubierre05b,zlochower05}. 

In \cite{hannam03} the puncture approach was extended to the CTS equations, and in \cite{hannam05} the conformal thin-sandwich puncture (CTSP) approach was applied to single boosted-black-hole initial-data sets. In this paper I present results for {\it binary}-black-hole initial-data sets. In particular, I locate quasi-circular orbits in these data and estimate the location of the ISCO. This work provides a useful comparison with excision data results, while avoiding altogether the issue of excision boundary conditions. It also makes it possible for evolution codes that employ punctures \cite{alcubierre03,brugmann04,imbiriba04,zlochower05} to take advantage of CTS data. This paper is an extension of the work discussed in \cite{hannam05}, which I will refer to as Paper I. 

There are currently two methods available to locate quasi-circular orbits in numerical initial-data sets. Cook's effective-potential method \cite{cook94} identifies orbits of particular angular momentum by locating minima in an effective potential with respect to the binary's separation, in analogy to Newtonian gravity. The mass-comparison method of Gourgoulhon, Grandcl\'{e}ment and Bonazzola \cite{ggba} assumes the existence of a helical Killing vector for black holes in quasi-circular orbit, and this manifests itself through the equality of the ADM and Komar mass estimates. Both the effective-potential \cite{cook94,baum00} and mass-comparison \cite{tichy04} methods have been applied to Bowen-York data, and are in good agreement in their predictions of quasi-circular orbits and the location of the ISCO. For CTS data all published work to date has used the mass-comparison method to locate orbits \cite{ggbb,yo04,cook04}.  

It was argued in \cite{hannam03} that it will be difficult (and perhaps impossible) to locate orbits in CTSP data using the mass-comparison method. However, this argument does not imply that orbits do not exist for CTSP data, and does not rule out using instead the effective-potential method to locate them. As such, in this paper I locate quasi-circular orbits in CTSP data using the effective-potential method.

In Section \ref{sec:background} I review the CTSP equations, and the procedure for constructing multiple boosted CTSP black holes. In Section \ref{sec:orbits} I review the effective-potential method. Section \ref{sec:methods} is devoted to numerical methods and the setup of the problem for equal-mass irrotational binary black holes. Results are presented in Section \ref{sec:results}.

\section{CTSP black-hole initial data}
\label{sec:background}

The conformal thin-sandwich (CTS) formulation proceeds from the 3+1 decomposition of Einstein's equations \cite{adm,york79,york98,york03}. In this decomposition spacetime is foliated into spacelike slices, and each slice is fully described by a spatial metric $\gamma_{ij}$ and an extrinsic curvature $K_{ij}$. The slices are connected via a shift vector $\beta^i$ and a lapse function $N$, allowing us to write the spacetime metric as \begin{equation}
ds^2 = -N^2 dt^2 + \gamma_{ij} (dx^i + \beta^i dt) ( dx^j + \beta^j dt).
\end{equation} The extrinsic curvature is given by \begin{equation}
K_{ij} = - \frac{1}{2N} \left( \partial_t \gamma_{ij} - \nabla_i \beta_j - \nabla_j \beta_i \right),
\end{equation} where the covariant derivative $\nabla_i$ is with respect to the spatial slice. It proves convenient to separate the extrinsic curvature into its trace and tracefree parts, $K_{ij} = A_{ij} + \frac{1}{3} \gamma_{ij} K$. 

In the CTS formulation, the physical quantities $\gamma_{ij}$, $A_{ij}$, $K$, $N$ and $\beta^i$, are all related to conformal background quantities via a conformal factor $\psi$: \begin{eqnarray}
\gamma_{ij} & = & \psi^4 \tilde{\gamma}_{ij}, \\
A_{ij} & = & \psi^{-2} \tilde{A}_{ij}, \\
K & = & \tilde{K}, \\
N & = & \psi^6 \tilde{N}, \\
\beta^i & = & \tilde{\beta}^i.
\end{eqnarray} For a good explanation of these conformal weightings, and how they relate to the conformal weightings in other decompositions, see \cite{york03}.

The extra ingredients in the conformal thin-sandwich {\it puncture} (CTSP) system are the puncture splittings of the conformal factor and lapse function. For $n$ black holes they are written as \begin{eqnarray}
\psi & = & 1 + \sum_i^n \frac{m_i}{2r_i} + u \label{eqn:psi} \\
\tilde{N} \psi^7 = N \psi & = & 1 + \sum_i^n \frac{c_i}{2r_i} + v. \label{eqn:lapsesplit} 
\end{eqnarray} The quantity $r_i$ is the coordinate distance from the $i$th puncture to the field point, $m_i$ parameterizes the mass of the $i$th black hole, and $c_i$ parameterizes the value of the lapse at the $i$th puncture. In the special case of a single stationary black hole, the solution $u = v = 0$ and choices $m_1 = M$ and $c_1 = -M$ provide us with the conformal factor and lapse of the Schwarzschild solution in isotropic coordinates. In nontrivial examples, the motivation for the puncture splittings is that they analytically remove the $1/r_i$ divergences in $\psi$ and $\tilde{N} \psi^7$ at the locations (``punctures'') of the black holes, and allow the initial-value equations to be solved numerically over the entire computational domain without any need for regions around the black holes to be excised. 

In any decomposition of the initial-value equations, there are variables that can be freely specified. In the CTS formulation the free data are the conformal metric, $\tilde{\gamma}_{ij}$, the trace of the extrinsic curvature, $K$, and their time derivatives, $\tilde{u}_{ij} \equiv \partial_t \tilde{\gamma}_{ij}$ and $\partial_t K$. The appearance of $\tilde{u}_{ij}$ and $\partial_t K$ in the free data is one of the advantages of the CTS formulation. Part of the dynamics of the spacetime can be freely specified, and this is ideal for constructing binary black holes in quasiequilibrium: we simply choose $\tilde{u}_{ij} = \partial_t K = 0$ \cite{cook00,cook04}. I will make these quasiequilibrium choices. For want of a better option, I also choose conformal flatness ($\tilde{\gamma}_{ij} = f_{ij}$, the flat metric) and maximal slicing, $K = 0$. 

With these choices, the CTSP initial-value equations in vacuum are \begin{eqnarray}
\tilde{\nabla}^2 u & = & - \frac{1}{8} \psi^{-7} \tilde{A}_{ij} \tilde{A}^{ij}, \label{eqn:CTSPHC} \\
\tilde{\Delta}_{\mathbb L} \beta^i - ( {\tilde{\mathbb L}} \beta )^{ij} \tilde{\nabla}_j \ln \tilde{N} & = & 0,  \label{eqn:CTSPMC} \\
\tilde{\nabla}^2 v & = & \tilde{N} \psi^7 \left[ \frac{7}{8} \psi^{-8} \tilde{A}_{ij} \tilde{A}^{ij}  \right]. \label{eqn:CTSPconstK} 
\end{eqnarray} Equation (\ref{eqn:CTSPHC}) is the Hamiltonian constraint, (\ref{eqn:CTSPMC}) is the momentum constraint, and (\ref{eqn:CTSPconstK}) is the maximal-slicing equation. The derivative operators are defined with respect to the conformal space, i.e., they are flat-space operators with the choice of conformal flatness. The longitudinal derivative and Laplacian are defined by \begin{eqnarray}
 ( \tilde{\mathbb L} \beta)^{ij} & \equiv & \tilde{\nabla}^i \beta^j + \tilde{\nabla}^j \beta^i - \frac{2}{3} \tilde{\gamma}^{ij} \tilde{\nabla}_k \beta^k, \\
\tilde{\Delta}_{\mathbb L} \beta^i & \equiv & \tilde{\nabla}^2 \beta^i + \frac{1}{3} \tilde{\nabla}^i \tilde{\nabla}_j \beta^j + \tilde{R}^i_j \beta^j,
\end{eqnarray} and conformal flatness gives $\tilde{R}^i_j = 0$

The conformal extrinsic curvature is constructed from the lapse and shift via \begin{equation}
\tilde{A}_{ij} = \frac{1}{2\tilde{N}} \left[ ( \tilde{\mathbb L} \beta )_{ij} - \tilde{u}_{ij} \right] ,
\end{equation} and the choice of $\tilde{u}_{ij} = 0$ reduces this to \begin{equation}
\tilde{A}_{ij} = \frac{1}{2\tilde{N}}  ( \tilde{\mathbb L} \beta )_{ij}. \label{eqn:Aij}
\end{equation}

When producing nontrivial solutions of the CTSP equations, it is necessary for the constants $c_i$ to be positive. If $c_i < 0$, the conformal lapse will pass through zero on some surface(s).  A calculation of the conformal extrinsic curvature using (\ref{eqn:Aij}) involves division by $\tilde{N}$, and division-by-zero errors will result if $\tilde{N}$ is allowed to pass through zero. It was shown in Paper I that different (positive) choices of $c_i$ have a negligible effect on the physical properties of single boosted-black-hole initial-data sets, and the effect of varying $c_i$ on binary-black-hole initial-data sets will be considered in Section \ref{sec:results}.

Note that the CTSP Hamiltonian and momentum constraints can be easily related back to the form of the constraints in the old CTT decomposition \cite{york03}. We may solve only (\ref{eqn:CTSPHC}) and (\ref{eqn:CTSPMC}), and choose $\tilde{N}$ instead of $\partial_t K$ as a piece of free data, removing the need to solve (\ref{eqn:CTSPconstK}). If we keep all of the free data choices above, but replace $\partial_t K = 0$ with $\tilde{N} = \tilde{N}_0$, a constant, there is an analytic solution of the momentum constraint for a boosted or spinning black hole: the Bowen-York solution. This solution can be used to construct the conformal extrinsic curvature for two boosted black holes, and has been used extensively in the past to construct binary-black-hole spacetimes. When viewed from the CTS viewpoint, we see that Bowen-York data can be considered to satisfy one quasiequilibrium requirement, $\partial_t \tilde{\gamma}_{ij} = 0$. The choice $\tilde{N} = \tilde{N}_0$, however, has no physical motivation. (Indeed, it implies an unattractive choice of lapse function, $N = \tilde{N} \psi^6 = \tilde{N}_0 \psi^6$, which diverges at the punctures.) The essential difference between Bowen-York data and full CTS (or CTSP) data is due to the addition of $\partial_t K$ to the free data, for which a choice can be physically motivated. 

For single boosted black holes, it seems that the additional requirement $\partial_t K = 0$ makes little difference to the physical content of the data (see Paper I). However, for binary black holes Tichy,  Br\"{u}gmann and Laguna have shown \cite{tichy03} that Bowen-York data do not simultaneously satisfy both $\tilde{u}_{ij} = 0$ and $\partial_t K = 0$. We therefore expect CTS and Bowen-York data to differ noticeably in the binary-black-hole case. 

Producing boosted-black-hole solutions to the full CTSP system (\ref{eqn:CTSPHC}), (\ref{eqn:CTSPMC}) and (\ref{eqn:CTSPconstK}) also requires a condition on the shift vector at each puncture. Boosted black holes can be generated by specifying a (non-zero) value of the shift vector at each puncture, as outlined in Paper I. This shift condition is of the form \begin{equation}
\beta^i = B_a^i
\end{equation} at the $a$th puncture, and results in a black hole boosted in the $B_a^i / | B_a^i |$ direction. The momentum of each black hole is parameterized by $B^i_a$, although that momentum is not known analytically and must be calculated numerically, as was done in Paper I. This procedure allows the construction of binary-black-hole initial data for irrotational black holes --- each black hole has non-zero linear momentum, but zero angular momentum, with respect to a distant observer. It is for equal-mass irrotational binary-black-hole spacetimes that I locate quasi-circular orbits and estimate the location of the ISCO in Section \ref{sec:results}. Before doing that, I will review two methods for locating quasi-circular orbits.

\section{Quasi-circular orbits}
\label{sec:orbits}

The first method developed to determine quasi-circular orbits in binary-black-hole initial-data sets was Cook's effective-potential method \cite{cook94}. The effective-potential method can be motivated by analogy with Newtonian gravity \cite{cook94,baum01}. One can locate circular orbits for a two-body system in Newtonian gravity by identifying, for a given value of the total angular momentum $J$ of the system, a minimum in the total energy $E$ of the system as a function of the separation $D$ of the two bodies. The criterion for a circular orbit is then \begin{equation}
\left. \frac{\partial E}{\partial D} \right|_J = 0.
\end{equation} The angular velocity of the binary system is given by \begin{equation}
\left. \Omega = \frac{\partial E}{\partial J} \right|_r.
\end{equation} This procedure leads to Kepler's law for Newtonian orbits. We can use a similar procedure in general relativity if we define an ``effective potential'' $E_b$ to take the place of the energy $E$ in the Newtonian case, and if we are careful to keep the individual masses of the black holes constant throughout the process. Cook \cite{cook94} defines an effective potential using the total ADM energy of the spacetime, \begin{equation}
E_b = E_{ADM} - M_1 - M_2. \label{eqn:ep}
\end{equation} In the original formulation of the effective-potential method, the individual black-hole masses $M_1$ and $M_2$ were defined in terms of the irreducible mass \cite{cook94,christodoulou70}. In later work Baker \cite{baker02b} used the puncture ADM masses of the black holes. Tichy and Br\"{u}gmann \cite{tichy04} found that, in the case of Bowen-York puncture data, the puncture-ADM and irreducible masses agree to within numerical error. In this work I have used the puncture ADM masses. 

We also need to choose a suitable measure of the black-hole separation, $D$. This can be the proper distance between the apparent horizons of the two black holes \cite{cook94,baum00}. If we are to measure the proper distance, we must first find the apparent horizon of each black hole. A simpler option is to use, as in this work, the coordinate separation between the two punctures \cite{baker02b,tichy04}. 

Given a binary-black-hole initial-data set, we can identify quasi-circular orbits by choosing some total angular momentum $J$, varying the separation of the black holes, $D$ (keeping $M_1$ and $M_2$ constant in the process), and searching for a minimum in $E_b$, i.e., \begin{equation}
\left. \frac{\partial E_b}{\partial D} \right|_{M,J} = 0.
\end{equation} This procedure can be repeated for different values of $J$. As $J$ decreases, the black-hole separation of the orbits also decreases, but at some point we will find that no orbit exists. The last orbit is identified as the innermost stable circular orbit (ISCO), which does not exist in Newtonian gravity. This method has been applied to both excision \cite{cook94} and puncture \cite{baum00,baker02b} Bowen-York data, and the ISCO predictions from both types of solution are in good agreement.

More recently Grandcl\'ement, Gourgoulhon and Bonazzola \cite{ggba,ggbb} developed an alternative method to locate quasi-circular orbits, which I will refer to as the mass-comparison method. The mass-comparison method assumes the existence of a helical Killing vector in a spacetime that contains two black holes in quasiequilibrium, which is realized through the equality of the ADM and Komar estimates of the total energy of the spacetime. The mass-comparison method has been applied to excision CTS data \cite{ggbb,cook04,yo04} and puncture Bowen-York data \cite{mdhPhD,tichy04}.

In the case of Bowen-York puncture data, the effective-potential and mass-comparison methods agree in their predictions of the parameters of quasi-circular orbits and the location of the ISCO \cite{tichy04}. The two methods also appear to agree well when applied to CTS excision data for corotational equal-mass binaries \cite{grigsby05}. Neither method has been applied to CTS-puncture data. 

As was mentioned in the Introduction, it may not be possible to apply the mass-comparison method to CTSP data. In particular, to locate orbits in the mass-comparison method one finds that $c_i < 0$ \cite{hannam03}, while numerical CTSP solutions require $c_i > 0$, as explained in Section \ref{sec:background}. I will instead use the effective-potential method. The disadvantage of the effective-potential method is that it does not determine a value for the lapse parameters $c_i$ in the decomposition (\ref{eqn:lapsesplit}). If the effective-potential method is to give meaningful results, the physical parameters of the quasi-circular orbits and ISCO will need to be shown to depend only weakly on different choices of the parameters $c_i$. I will discuss this point further in Section \ref{sec:results}.

\section{Numerical setup}
\label{sec:methods}

I consider equal-mass black-hole binary systems, where $m_1 = m_2 = m$ and $c_1 = c_2 = c$ in (\ref{eqn:psi}) and (\ref{eqn:lapsesplit}). The CTSP equations (\ref{eqn:CTSPHC}) -- (\ref{eqn:CTSPconstK}) are solved numerically on a Cartesian grid, using a multigrid solver that was adapted from the {\tt BAM\_Elliptic} solver in the Cactus infrastructure \cite{cactus,hannam05}. In Paper I this solver was found to be between first- and second-order convergent when solving the CTSP equations for boosted black holes. 

The punctures are located on the $z$-axis at $z = \pm D/2$, where $D$ is the coordinate distance between the two punctures. Each black hole is boosted in the $x$-direction by specifying \begin{equation}
\beta^i = (\mp B, 0, 0) \label{eqn:shift_condn}
\end{equation} at its respective puncture. In the single-black-hole results presented in Paper I, the puncture was located at the origin; condition (\ref{eqn:shift_condn}) could be imposed on every level of the multigrid hierarchy because the origin existed on every level. In the binary-black-hole case the puncture is placed on a point that exists on the finest grid, but may not exist on any of the coarser subgrids. Condition (\ref{eqn:shift_condn}) is applied on the finest grid, and on each of the coarser grids the value of the shift vector is specified at one of the points nearest to the puncture, and given a value such that first-order interpolation would give the correct value ($\mp B$) at the puncture.

The CTSP equations are solved on only one octant of the Cartesian grid (i.e., for only positive $x$, $y$ and $z$), by using the symmetries of the CTSP variables consistent with a binary system with non-zero angular momentum, shown in Table \ref{tab:symmetries}. These symmetries are found by noting that as $r \rightarrow \infty$, the CTS momentum constraint (\ref{eqn:CTSPMC}) approaches the form of the CTT momentum constraint, \begin{equation}
\tilde{\Delta}_{\mathbb L} \beta^i = 0,
\end{equation} for which we have the Bowen-York solutions for a boosted or spinning black hole. Far from the binary, I assume that the solution of the momentum constraint will approach that for a single spinning black hole \cite{omurch92}, and that the shift vector has the angular and radial dependence of the vector potential for a single spinning Bowen-York black hole, \begin{equation}
\beta^i \sim \frac{\epsilon^{ijk} n_j \hat{J}_k}{r^2}, \label{eqn:ob_shift}
\end{equation} where $\hat{J}^i$ is the direction of the total angular momentum of the system and $n^i$ is the radial normal vector directed away from the origin. The same symmetries were used by Baumgarte, {\it et. al.} \cite{baum97b} to solve the conformal thin-sandwich equations for binary neutron stars. For the numerical setup I have described, $\hat{J}^i = (0,1,0)$. 

In the multigrid solver, careful treatment of the symmetry planes was required whenever a component of the shift vector was odd across that plane. The odd symmetry requirement is that the function has equal and opposite values on either side of the symmetry plane, and be zero on the plane. However, at intermediate steps in the solution procedure the function is unlikely to be exactly zero at the points on the symmetry plane. I have found that an improvement in efficiency of the code by many orders of magnitude can be achieved by setting the function to zero on the symmetry plane before any prolongation operation (which interpolates the function from a coarse to a fine grid), but not at any other time. 

At the outer boundary, the functions $u$ and $v$ fall off as $1/r$. The behavior of the shift vector at the outer boundary is required to be consistent with (\ref{eqn:ob_shift}). For the setup described, the condition (\ref{eqn:ob_shift}) applies to only two components of the shift vector, $\beta^x$ and $\beta^z$; the $1/r^2$ contribution to $\beta^y$ is zero. As such, the code requires that $\beta^y$ fall off as $1/r^3$. The outer boundary behavior of the CTSP variables is shown in Table \ref{tab:symmetries}.

\begin{table}
\begin{center}
\begin{tabular}{ccccc} 
          & $x = 0$ & $y = 0$ & $z = 0$ &  $r \rightarrow \infty$ \\ \hline
$u$       & even    &  even   &  even   &  $1/r$ \\
$v$       & even    &  even   &  even   &  $1/r$ \\
$\beta^x$ & even    &  even   &  odd    &  $z/r^3$ \\
$\beta^y$ & odd     &  odd    &  odd    &  $1/r^3$ \\
$\beta^z$ &  odd    &  even   &  even   &  $x/r^3$ \\
\end{tabular}
\caption{Coordinate plane symmetries and outer boundary conditions for CTSP variables for an equal-mass irrotational binary-black-hole system with angular momentum in the $y$-direction.} \label{tab:symmetries}
\end{center}
\end{table}

To enforce the asymptotic behavior shown in Table \ref{tab:symmetries}, the code applies Robin outer-boundary conditions to all of the CTSP variables. The conditions applied to $u$ and $v$ are \begin{eqnarray}
N^i \partial_i (ru) & = & 0, \label{eqn:robin_u} \\
N^i \partial_i (rv) & = & 0, \label{eqn:robin_v}
\end{eqnarray} where $N^i$ is the unit vector perpendicular to the boundary. The conditions applied to the components of the shift vector are \begin{equation}
N^l \partial_l (r^2 \beta^i) = \frac{1}{r} \epsilon^{ijk}  ( N_j - n_j n_l N^l) \hat{J}_k. \label{eqn:robin_shift}
\end{equation} for $\beta^x$ and $\beta^z$, and \begin{equation}
N^i \partial_i (r^3 \beta^y) = 0 \label{eqn:robin_betay}
\end{equation} for $\beta^y$.

For a given solution of the CTSP equations with the shift condition (\ref{eqn:shift_condn}) and outer boundary conditions (\ref{eqn:robin_u}), (\ref{eqn:robin_v}), (\ref{eqn:robin_shift}) and (\ref{eqn:robin_betay}), we can calculate a number of physical quantities. The total angular momentum of the spacetime is given by \cite{bowen80,cook94} \begin{equation}
J_i = \frac{\epsilon_{ijk}}{8 \pi} \oint x^j \tilde{A}^{kl} d^2 S_l. \label{eqn:jint}
\end{equation} The surface integral is computed at the outer boundary of the computational grid. The integral was constructed using global Killing vectors of the conformal space (which are asymptotic Killing vectors of the physical space) and can in fact be computed at any radius, so long as the surface surrounds the puncture. For the setup we have described, the symmetries will enforce $J^x = J^z = 0$. 

The total ADM mass of the spacetime is given by \cite{omurch74c,baum00} \begin{equation}
M_{ADM} = \sum_i^n m_i + \frac{1}{16\pi} \int \psi^{-7} \tilde{A}_{ij} \tilde{A}^{ij} dV.
\label{eqn:ADMintsplit}
\end{equation} The volume integral in (\ref{eqn:ADMintsplit}) is evaluated over the entire numerical grid. The contribution from outside the numerical grid is estimated with the aid of the outer boundary conditions, using a technique described in Paper I. 

There is no unique measure of the mass of each individual black hole. I have chosen to use the ADM mass computed on the second hypersurface of each black hole, sometimes called the bare mass \cite{brill63,brandt97}. I will refer to it as the puncture ADM mass, because it can be easily computed at each black hole's puncture. The puncture ADM mass, $M_i$, is \cite{brandt97} \begin{equation}
M_1 = m_1 \left( 1 + u_1 + \frac{m_2}{2D} \right), \label{eqn:bare_mass}
\end{equation} where $u_1$ is the value of the function $u$ at the first puncture, and $M_2$ is found by interchanging the subscripts ``1'' and ``2''. Note that due to the symmetries of the equal-mass binary case, $M_1 = M_2$. Define the total black-hole mass as $M = M_1 + M_2$ and the reduced mass as $\mu = M_1 M_2 / (M_1 + M_2)$, which are $M = 2$ and $\mu = 1/2$ in the results presented below. 

When producing lines of constant $J$ and $M$ in the effective-potential method, the constants $B$ in (\ref{eqn:shift_condn}) and $m$ in (\ref{eqn:psi}) are varied until $J$ and $M$ are equal to the required values, with a combined error of less than $3 \times 10^{-6}$.

\section{Results}
\label{sec:results}

In the first set of results presented here, the lapse parameter is $c = 1.0$. Quasi-circular orbits were found for numerical solutions with a resolution of $h = 0.0625M$ and an outer boundary at $16M$. The innermost stable circular orbit was identified at $\bar{J} = J/(\mu M) = 3.115 \pm 0.005$, $\bar{D} = D / 2M = 1.625 \pm 0.125$, and $\bar{E}_b = E_b / \mu = -0.0775 \pm 0.002$. The orbital angular speed at the ISCO was $\bar{\Omega} = M\Omega = 0.11 \pm 0.01$. 

The uncertainties in these quantities were found by locating the ISCO for solutions with the same resolution but a more distant outer boundary ($h = 0.0625M$, outer boundary at $24M$). There is also uncertainty due to the freedom in the lapse parameter $c$. As discussed in Section \ref{sec:orbits}, the lapse parameter $c$ is undetermined by the effective-potential method. I investigated the dependence of the location of the ISCO on this parameter for the values $c = 0.25, 0.5, 1.0$, and $2.0$. Note that the elliptic CTSP system becomes more difficult to solve as $c$ decreases, and I could not find solutions for $c = 0$. This is unfortunate, because when $c=0$ the lapse would be zero at the black-hole punctures, and constitute a fully ``pre-collapsed lapse'', a potentially desirable choice for numerical evolutions \cite{alcubierre03,alcubierre05,reimann04a,reimann04b}.

\begin{table}
\begin{center}
\begin{tabular}{cccccc} 
$c$  &  $\bar{D}$ & $\bar{J}$ & $\bar{E}_b$ &  $E_{ADM}$ & $\bar{\Omega}$ \\ \hline
0.25 &   1.625    & 3.1175    &  -0.0774    &  1.961326  &    0.1144      \\
0.5  &   1.625    & 3.1175    &  -0.0773    &  1.961342  &    0.1137      \\
1.0  &   1.625    & 3.1150    &  -0.0775    &  1.961228  &    0.1118      \\
2.0  &   1.625    & 3.1125    &  -0.0777    &  1.961169  &    0.1133      \\
\end{tabular}
\caption{ISCO results for various values of the lapse parameter $c$, with resolution $h = 0.125m$ and outer boundary at $32m$.} \label{tab:isco_c}
\end{center}
\end{table}

The ISCO parameters for different choices of $c$ are shown in Table \ref{tab:isco_c}. It is clear that the value of $c$ has only a small effect on the ISCO location, and the values remain well within the uncertainties quoted above. 

\begin{figure}[!ht]
\begin{center}
\includegraphics[scale=0.41]{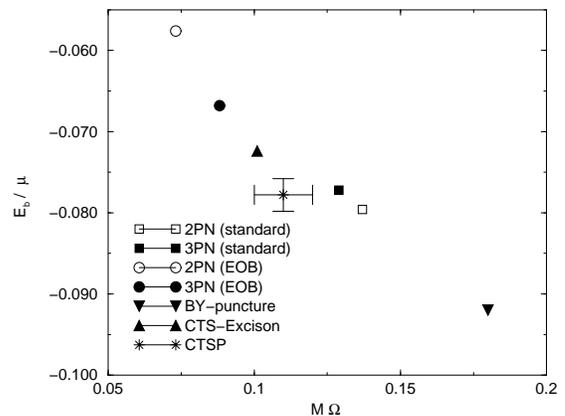}
\caption{ISCO results for equal-mass irrotational binaries. The methods are as described in Table \ref{tab:comparison}. The error bars for the CTSP result are based on the uncertainties quoted in the text, with $c = 1.0$.}
\label{fig:isco}
\end{center}
\end{figure}

Table \ref{tab:comparison} compares ISCO results for irrotational binaries, obtained by different methods. The ``BY-Puncture'' results at the end of the table are those obtained by Baumgarte \cite{baum00} for Bowen-York puncture data. These results are not close to any of the post-Newtonian results, from either effective one-body (EOB) \cite{damour02} or standard post-Newtonian \cite{blanchet02} methods. The CTSP results in this paper and the CTS excision results of Cook and Pfeiffer \cite{cook04} agree in their estimate of $\bar{\Omega}$ at the ISCO. They do not agree on $\bar{J}$ and $\bar{E}_b$, but are very close. Both sets of results are much closer to the post-Newtonian estimates than the Bowen-York-puncture results --- in fact, they are between the 2PN and 3PN results from the ``standard'' and effective one-body methods. The various ISCO values of $\bar{E}_b$ and $\bar{\Omega}$ from different methods are also shown in Figure \ref{fig:isco}. 

\begin{table}
\begin{center}
\begin{tabular}{lcccc}
Method         &  $\bar{J}$   &   $\bar{E}_b$   &   $\bar{\Omega}$ \\ \hline
CTSP           &    3.115     &     -0.0775     &     0.112        \\
CTS-Excision   &    3.06      &     -0.0724     &     0.101        \\ \hline
2PN (standard) &    3.116     &     -0.0796     &     0.137        \\
3PN (standard) &    3.144     &     -0.0772     &     0.129        \\ \hline
2PN (EOB)      &    3.408     &     -0.0576     &     0.0732       \\
3PN (EOB)      &    3.28      &     -0.0668     &     0.0882       \\ \hline
BY-Puncture    &    2.95      &     -0.092      &     0.18         \\
\end{tabular}
\caption{ISCO results for irrotational equal-mass binaries from different methods. The CTS-Excision results are typical values from Cook and Pfeiffer \cite{cook04}. The ``PN (standard)'' results are from Blanchet \cite{blanchet02} using a standard post-Newtonian expansion. The ``EOB'' results are from effective one-body post-Newtonian methods \cite{damour02}. The BY-puncture results are for Bowen-York puncture data \cite{baum00}.} \label{tab:comparison}
\end{center}
\end{table}

\section{Discussion}

I have numerically solved the CTSP equations to construct initial-data sets for irrotational binary-black-hole spacetimes. Contrary to the implications of \cite{hannam03}, it is possible to locate quasi-circular orbits in CTSP data, by using the effective-potential method instead of the mass-comparison method. Identification of quasi-circular orbits allows an estimate of the innermost stable circular orbit (ISCO). The physical parameters of the ISCO are close to those found by Cook and Pfeiffer \cite{cook04}, who solved the CTS equations with excision techniques and used the mass-comparison method to locate quasi-circular orbits. This suggests a great deal of robustness of the results: essentially the same CTS ISCO was found using two different solution procedures (punctures and excision), and two different methods of determining quasi-circular orbits (the effective-potential and mass-comparison methods). These results also suggest that, as in the case of Bowen-York data \cite{tichy04}, the effective-potential and mass-comparison methods agree in their determination of quasi-circular orbits.

Although the mass-comparison method was not used to locate quasi-circular orbits in the CTSP data, it is natural to wonder how close the ADM and Komar mass estimates are for these data. Recall that in the mass-comparison method equality of the two mass estimates is considered to denote a quasi-equilibrium spacetime \cite{ggba}. One immediately finds that in the CTSP approach, the ADM and Komar masses cannot be equal. This can be seen by writing out the Komar mass as was done for the ADM mass in (\ref{eqn:ADMintsplit}). The ADM mass for an equal-mass binary can be written \begin{equation} 
M_{ADM} = 2 m + I_1,
\end{equation} where $I_1$ is the integral in (\ref{eqn:ADMintsplit}), and $m = m_1 = m_2$ is the mass parameter. The Komar mass can be written \cite{tichy03,mdhPhD} \begin{equation}
M_K = m - c + I_1/2 + I_2,
\end{equation} where $I_2$ is similar to $I_1$. In particular, $I_1$ and $I_2$ are both positive, and small compared to the mass parameter $m$. If $M_{ADM}$ and $M_K$ are to agree, we need roughly $c = -m$. Using $c > 0$, as in the CTSP procedure, $M_{ADM}$ and $M_K$ will always disagree. They will be closest for the smallest choice of $c$ (which is 0.25 in the results reported here), but still differ by more than a factor of two. 

The results suggest that the physical content of the data changes little as $c$ is changed, while the Komar mass will change a great deal. One could argue that the data with the closest ADM and Komar masses (i.e., when $c$ is smallest) represent the data closest to quasiequilibrium. However, it is not clear how one could quantify what "close" means in this case. 

It is now possible to construct CTS binary-black-hole initial data for use in codes that use punctures. In puncture evolutions the lapse quickly collapses to zero at the punctures \cite{alcubierre03,zlochower05}, and codes can become unstable if the lapse is below zero anywhere on the numerical grid. CTSP data are well suited to both of these requirements. For suitable choices of the parameters $c_i$, the lapse is partially ``pre-collapsed'' at the punctures, and the lapse by construction must always be positive. Evolution of these data sets will be the subject of further research.

\acknowledgments

I thank Greg Cook, Manuela Campanelli, Bernard Kelly, Steve Lau, Carlos Lousto and Yosef Zlochower for useful discussions. I gratefully acknowledge the support of the NASA Center for Gravitational Wave Astronomy at The University of Texas at Brownsville (NAG5-13396) and from NSF grants PHY-0140326 and PHY-0354867. All numerical results were obtained on the CGWA Funes cluster.

\bibliography{ctsp}

\end{document}